\documentclass{cpcauth}
\usepackage{psfig}
\begin{document}
\begin{frontmatter}
\title{On the distribution function of the information speed 
\\ in computer network}
\author[CH,NA]{Lev N. Shchur\thanksref{em}}
\address[CH]{Landau Institute for Theoretical Physics, 142432
Chernogolovka, Russia}
\address[NA]{Laboratoire de Physique des Mat\'eriaux$^{\, 2}$,
Universit\'e Henri Poincar\'e Nancy I, B.P. 239,
F--54506 Vand{\oe}uvre l\`es Nancy Cedex, France}
\thanks[em]{e-mail: lev@landau.ac.ru}
\thanks[la]{Laboratoire associ\'e au CNRS UMR 7556}

\begin{abstract}

We review a study of the Internet traffic properties. We analyze under
what conditions the reported results could be reproduced. Relations of
results of passive measurements and those of modelling are also discussed.
An example of the first-order phase transitions in the Internet traffic is
presented.

\end{abstract}
\end{frontmatter}


\section{Introduction}

For the Internet user the most important parameter of the network is the
speed at which one retrieves documents.  Anyone browsing the Internet for
preprints and articles (xxx.lanl.gov, publish.aps.org, elsevier.nl,
iop.org, wspc.com, etc.), or looking for the news and weather, booking tickets,
making hotel reservations, etc. asks himself: {\em why is the Web so
slow?} This is precisely the question which motivates our work.

The Internet efficiency depends upon the two main aspects of the network,
topology and transport. The topology and connectivity features of the
Internet were discussed by Newmann and Barabasi at this
Conference~\cite{N01,B01}, and our main subject is the review of the
Internet transport investigation: its measurements, properties and
modelling.

The first set of properties is connected with natural characteristics of
human activity. In this connection, the daily working hours lead to a
daily periodicity of the web traffic, the work week leads to a weekly
periodicity (weekends!), the annual calendar leads to an annual
periodicity (winter and summer vacations!), and so on. Holidays and
important events (political campaigns, etc.) can also affect the traffic.
The latest and most sudden example is the congestion of all news servers
on September 11 of 2001 just after the terrorist attack on America.

The second set of the web traffic properties is connected with the fact
that the path from one point on the web (e.g., user) and another point
(e.g., server) is not stable. The path is often quite complicated and
consists of a number of routers, channels, caches, etc., which changes
with time because the network constantly develops~\cite{KS01,KS01b}. This
reminds us of the ancient philosopher Heraclites, who asserted ``You
cannot step twice into the same river''. We could say the same about the
Internet river. The third set of properties further complicating the
measurements is connected with the dynamics of the Internet traffic for
many autonomous systems, which leads to the random changes in the
transport topology and therefore in the timing and loading
characteristics.

The knowledge obtained on the Internet traffic properties is reviewed in
this article.

\section{Network traffic: $1/f$ noise and models.}

Network traffic has been the subject of intensive investigations
especially in the last eight years after WWW technology was invented.
Nevertherless, we still do not have a single simple answer to the main
question of {\em why is the Web so slow?}.

One of the main difficulties of understanding the Internet nature is that
it is neither centrally planned nor configured. It is a good example of
the self-organised global structure which is still a self-growing one.
People who are running the Internet started in 2000 their annual workshop
on the passive and active measurements (PAM) on the Internet. The relation
between passive and active measurements is still not understood. Active
measurements are easily understood but do not clearly predict actual
Internet and Web performance; passive Internet/Web measurements can
reflect actual performance but can be hard to interpret. Passive
measurement is nothing but data acquisition to log-files of the
transactions routinely done by the routers, switches, proxies, caches and
workstations. One could find this information in the corresponding
log-files. The active measurements could be divided into several groups.
The two most extreme are RTT information and modelling of the user
activity. RTT (round trip time) between two nodes on the Internet could be
obtained, f.e., using the usual Unix command {\em ping}. The resulting
value can be used by many protocols for different reasons like, f.e., for
a path optimization or more often just to check whether the given node is
accessible or not.

The distribution of RTT times is not trivial and it is the subject of
intensive investigation after the pioneering paper of Csabai~\cite{C94}.
Analysing the results of several hundreds of RTTs obtained in two weeks
between his workstation in Budapest, Hungary, and an ftp server in
Helsinki, Finland Csabai found that the power spectral density could be
described by power law $1/f^{1.15}$ in a wide range of frequencies $f$
from $10^{-4}\; Hz$ to $0.5\; Hz$.

Actually, the self-similarity in computer traffic was found a little bit
earlier by Leland et al.~\cite{LTWW94} for the packet flow density in
several local Ethernet networks at the Bellcore Morristown Research and
Engineering Center. They found ``heavy tails'' in the cumulative
distribution function of the packet sizes. At that time none of the
commonly used traffic models was able to capture this fractal-like
behavior.

Takaysu et al.~\cite{TTS96} developed a {\em contact process} (CP) model
of jam dynamics on the Internet. They associated a particle with the
non-jamming gateway which can reproduce another particle at a neighboring
site at a rate $p$, and can annihilate (i.e. jammed gateway) spontaneously
at a rate $q$, and $p+q \le 1$. They found that the survival probability
at the critical point ($\delta=1-p/q\rightarrow 0$) was proportional to
the inverse time $1/t$. Moreover, assuming that the RTT times are the
two-valued function, taking values $h$ and $0$, they showed that the power
spectrum proportional to $1/f^\alpha$ with value $\alpha$ bounded,
$0<\alpha \le 1$. This result seems to be supported by the analysis of
Ethernet and Internet traffic in a series of papers by Takayusu, et
al.~\cite{T-ser}.

Huisinga et al.~\cite{HBKSS} introduced a microscopic model for the packet
transport on the Internet. Data are divided into small packets of a
definite size. These data packets move, for fixed source and destination
hosts, due to the structure of TCP/IP, along a temporally fixed route.
Therefore, the transport between two specific hosts can be viewed as a
one-dimensional process. The cellular automaton model Assymetric Simple
Exclusion Process (ASEP) has an important property - the occurence of
boundary-induced pase transitions~\cite{K91}.

Analysing the power spectrum of the travel times Huisinga et
al.~\cite{HBKSS} identified three phases: free flow characterized by the
white noise, congested phase with $1/f^{1/2}$-noise at low frequencies and
white noise at high frequencies, and critical load phase with with
$1/f$-noise at low frequencies and white noise at high frequencies. They
concluded with the important observation that the jamming properties are
not related to the structure of the network and rather connected with the
paths with critical load.

\section{Server load and latency times}

Although the traffic models discussed in the previous section seem to
explain some pecularities of the Internet traffic at the critical path
load, some recently observed phenomena connected with the server critical
load (news servers!) are still not understood.

Barford and Crovella~\cite{BC99,BC01} find surprising effect of server
load. When the network is heavily loaded (i.e., packet loss rates are
high), it is not uncommon for a heavily loaded server to show a {\em
better} mean response time than a lightly loaded server. Their
measurements suggest that this may be because heavily loaded servers often
show lower packet loss rates, and since packet losses have dramatic
effects on transfer latency, this can reduce the mean response time.

In the rest of this section I will analyse the probability distribution
function of the latency times and arrive at the simplest experimental
setup having the property observed by Barford and Crovella
~\cite{BC99,BC01}.

It is a commonly accepted picture (see, f.e., Figure~1 in the paper by
Helbing, et al.~\cite{HHM}) that the distribution of download times (i.e.,
latency times) is log-normal and that this property leads to the $1/f$
property of the Internet traffic.

In fact, the distribution function of latency times is
multipeaked~\cite{KS01}. The peaks could be associated with two sets of
factors. The first ones are connected with the different throughput of the
particular paths between user worskstation and destination servers. This
fact is clearly visible for some preprint and reprint library servers
placed in the Far East. The next set of factors is connected with the
traffic content. The speed of document retrieval depends on the type of
document: text, image, binary file, audio, video, etc. Indeed, it is
practically possible to decompose the distribution function into the more
elementary ones analysing the content of the proxy server log files and
using the above-mentioned two sets of factors. Nevertherless, even in this
case, the distribution function of elementary processes like taking files
from a particular archive (say, Los Alamos Archive) to a given
workstation, the distribution more often demonstrates two
peaks~\cite{SK01}.

\begin{figure}
\center
\psfig{figure=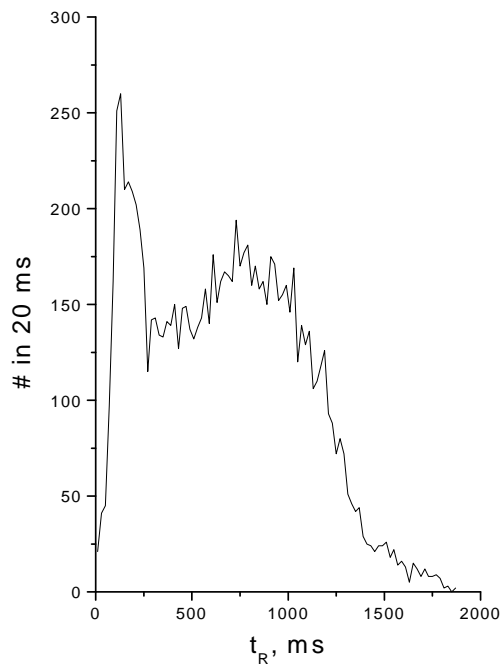,width=7.5cm,angle=0}
\caption{Typical histogram of RTT times (in $ms$) between workstation and 
router BNS045 calculated as number of RTT time values within an interval
of 20 $ms$.} 
\label{bns045}
\end{figure}

The same effect of multipeaked distributions could also be obtained
analysing RTT times on a short path between workstation and border router.
Figure~\ref{bns045} shows a histogram of RTT times between workstation
(connected to 100 Mbps/s Ethernet fiber-optic campus network of
Chernogolovka Science Park (AS~9113)) and border Cisco router BNS045
(147.45.20.221) of FREEnet. AS~9113 and BNS045 connected by a 2 Mbps ATM
channel. Workstation and border router are separated by only one LAN router.
The large and narrow peak at about 7 $ms$ is associated with the round
trip time of a 64 bit ping packet in the path connecting three devices
with an empty 2 Mbps ATM channel. The next and wide peak at about 750 $ms$
could be associated with the router congestions. The fact that these peaks
are well separated is due to some particular features of the TCP protocol.
This picture is a clear demonstration of the nonlinear response of the
router.

\begin{figure}
\center
\psfig{figure=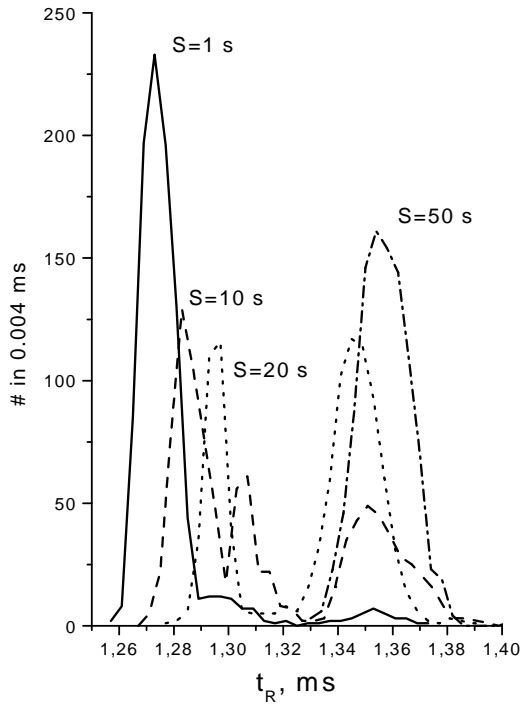,width=8cm,angle=0}  

\caption{Histograms of internal ping RTT times calculated in intervals of
$0.004$ $ms$ for the interval times between two ping packets $S=1$ $s$
(solid line), $S=10$ $s$ (dashed line), $S=20$ $s$ (dotted line), and
$S=50$ $s$ (dash-dotted line).}

\label{f3}
\end{figure}

 Moreover, we found~\cite{SK01} that RTT times inside a single
workstation exhibit usually two peaks in the probability distribution
function of RTT times. We performed an analysis of the RTT times of the
Unix ping command on the internal network interface

\begin{verbatim}

ping -i S -s 56 127.0.0.1

\end{verbatim}

\noindent where $S$ is the interval in seconds between two consecutive
ping packets of the size of 56 bytes (the ping packet contains also 8
additional bytes, i.e. the total packet size is 64 bytes). We vary interval
$S$ and accumulate data for $S=1$, $10$, $20$ and $50$ seconds.
Figure~\ref{f3} shows histograms $P(t_R)$ of RTT times calculated in
intervals of $0.004$ $ms$. This histograms were obtained from the results
of 50000 pings grouped in 5 groups and then averaged. Fluctuations from
one group to another are small enough.

All histograms have two peaks, placed at about $t_R^{(1)}=1.28-1.29$ $ms$
and $t_R^{(2)}=1.35-1.36$ $ms$. For the pings with interval $S=1$ $s$
there are higher probabilities that measured RTT time will be about
$t_R^{(1)}$. For the pings with interval $S=50$ $s$ this probability is
higher at the value of about $t_R^{(2)}$. For the intermediate interval
between pings $S=20$ $s$ both peaks have approximately the same height,
and the result of the measurement of $t_R^{(1)}$ or $t_R^{(2)}$ will be
equally probable.

\begin{figure}
\center
\psfig{figure=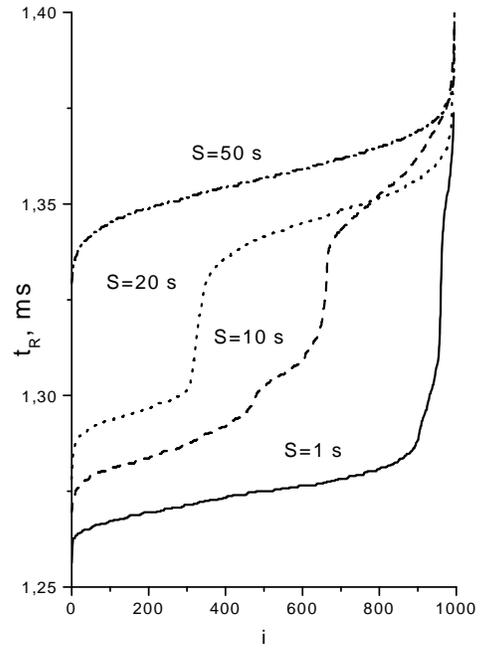,width=7.5cm,angle=0}

\caption{Internal ping RTT times ranked ascendingly for the interval times
between two ping packets $S=1$ $s$ (solid line), $S=10$ $s$ (dashed
line), $S=20$ $s$ (dotted line), and $S=50$ $s$ (dash-dotted line).} 

\label{t3}
\end{figure}

It is better to plot RTT times ranked ascendingly as shown on
Figure~\ref{t3}. Changing the axes and their directions one could find
that this curve after normalization is nothing but a cumulative
distribution function of RTT times. In fact, the curves in Figure~\ref{f3}
could be obtained from Figure~\ref{t3} with proper differentiation.
Figures~\ref{f3}-\ref{t3} clearly demonstrate that by varying the interval
between pings we have some kind of ``first-order phase transition''
between states characterised by RTT time values $t_R^{(1)}$ and
$t_R^{(2)}$. It is not a true phase transition because we could not
associate any order parameter with the process.

The effect we found is stable and reproducible. We obtain the same behaviour
for the number of Unix workstations of different types disconnected from
the network and not running any processes except minimal configurations.
The only difference is the values of $t_R^{(1)}$ and $t_R^{(2)}$ RTT times
and the interval $S$ times. So, this effect could be considered as the
universal one.

We associate this effect with the cache memory organization. It seems that
the difference between the values of $t_R^{(2)}$ and $t_R^{(1)}$ is
the time necessary to upload the ping process to the cache memory. Unix
systems were running some processes which could oust ping procedure from
the cache and thus increase the RTT time. A detailed analysis will be
published elsewhere~\cite{SK01} as well the the model of the process.

We analysed the power spectrum of the RTT signals~\cite{C94} and found
that in all cases it could be approximated with the $1/f^\alpha$ law with
$\alpha=2$ for the low frequencies and demonstrated white noise at the
high frequencies. Nevertheless, this analysis has to be considered with
more caution; we found in one measurement that just one fluctuation in RTT
time which gives $t_R=106$ $ms$ obtained in any experiment with $S=50$ $s$
has changed our power spectrum drastically. Excluding this enormous
fluctuation all results are very stable and reproducible.

\section{Discussion}

We have to note that the effect of first-order phase transition which we
found to be the influence of the cache memory, could be even more
universal. In fact, all servers usually use cache memory, and the effect
of the higher productivity of the servers~\cite{BC99,BC01} under the heavy
load could be explained as an effect of the heavy cache memory usage as
well.

Most of the models of Internet traffic are based on the assumption that
the routers are nothing but queues. It seems that this is not always the
case and more sophisticated nonlinear models of the elements of the
network should be considered. Most of the nodes which are just single
routers nowdays consist of several devices which separately work as
switches, or routers, or name servers, or proxy-cache servers with rather
complicated intercommunications and interactions.

\ack Thanks are due to Serge Krashakov and Timofey Rostunov for help and
discussions. This work is supported by the twin research program of the
Landau Institute and the Ecole Normale Sup\'erieure de Paris and by RFBR
under projects 99-07-18412, 99-07-90084, and 01-07-90119.

\end{document}